\documentstyle[aas2pp4]{article}

\newcommand\Ha{H$\alpha$}

\newcommand\hii{\ion{H}{2}}

\newcommand\etal{et\thinspace al.~}
\newcommand\eg{e.g.,~}

\newcommand\msol{\rm\,M_\odot}
\newcommand\ergs{{\rm\,erg\,s^{-1}}}

\newcommand\hlf{\hbox{H{\thinspace}{\small II}} LF}
\newcommand\lsat{L_{\rm sat}}
\newcommand\lup{l_{\rm up}}
\newcommand\mup{m_{\rm up}}

\received{16 September 1997}
\accepted{10 December 1997}

\slugcomment{Accepted 10-Dec-97 to the Astronomical Journal}

\lefthead{M. S. Oey and C. J. Clarke}
\righthead{\hii\ region luminosity function}

\begin{document}

\title{	On the form of the H\,{\Large\bf II} region luminosity function \\
 \medskip \rm\small Accepted 10-Dec-97 to the Astronomical Journal}

\author{M. S. Oey and C. J. Clarke}
\affil{Institute of Astronomy, Madingley Road, Cambridge, CB3 0HA,
	U.K. \\ oey@ast.cam.ac.uk; cclarke@ast.cam.ac.uk }


\begin{abstract}
Observed variations in the \hii\ region luminosity function (\hlf) seen
in spiral arm vs. interarm regions, and different galactic Hubble
type, can be explained simply by evolutionary effects and maximum
number of ionizing stars per cluster.
We present Monte Carlo simulations of the \hlf, drawing the number of
ionizing stars $N_*$ from a 
power-law distribution of constant slope, and the stellar masses from a 
Salpeter IMF with an upper-mass limit of $100\msol$.  We
investigate the evolution of the \hlf, as determined by stellar
main-sequence lifetimes and ionizing luminosities, for a single burst
case and continuous creation of the nebular population.  Shallower
\hlf\ slopes measured for the arms 
of spiral galaxies can be explained as a composite slope, expected for
a zero-age burst population, whereas the interarm regions tend to be
dominated by evolved rich clusters described by a single, steeper
slope.  Steeper slopes in earlier-type galaxies can be explained
simply by a lower maximum $N_*$ cutoff found for the parent OB associations.
The form of the \hlf\ can reveal features of the most recent
($\lesssim 10$ Myr) star formation history in nearby galaxies.
\end{abstract}


\keywords{\hii\ regions --- galaxies: fundamental parameters ---
galaxies: ISM --- galaxies: star clusters --- ISM: structure ---
stars: formation 
}

\section{Introduction}

\hii\ regions have long served as a primary indicator of
high-mass star formation, both in the Milky Way and in external
galaxies.  The luminosity function of \hii\ regions (\hlf) 
is therefore a vital probe of the present global star
formation properties in individual galaxies.  
In recent years, the number and accuracy of compiled
\hlf s for nearby galaxies has increased substantially, extending
the use of this tool to greater distances.

The differential \hlf\ is usually parameterized as a power law:
\begin{equation}
N(L)\ dL = A\ L^{-a}\ dL \quad ,
\end{equation}
where $N(L)\ dL$ is the number of nebulae with luminosities in the range
$L$ to $L + dL$.  Some interesting patterns have emerged regarding the
form of the \hlf: 
\begin{enumerate}
\item[1.]
A large number of galaxies show a break in slope of the \hlf\,
with the fainter \hii\ regions showing a shallower slope compared to
the high-luminosity objects (\eg Kennicutt, Edgar, \& Hodge 1989,
hereafter KEH; Rand 1992; Walterbos \& Braun 1992; Rozas, Beckman, \&
Knapen 1996a;)

\item[2.]
The \hlf\ for arm and interarm regions of spiral galaxies
sometimes show steeper slopes in the interarm regions (e.g., KEH;
Banfi {\etal}1993; Rand 1992).  However, there are also many
instances where no difference is found (e.g., Rozas
{\etal}1996a; Knapen {\etal}1993; Knapen 1997)

\item[3.]
The slope of the \hlf\ is correlated with galactic Hubble type, such
that early-type galaxies show steeper slopes than late-types.  The
power-law index $a \sim 2.0$ for Sb -- Sc galaxies, compared to $a\sim
1.7$ for Sc -- Im galaxies (KEH; Banfi {\etal}1993).  It is even
steeper in Sa galaxies, with $a\sim 2.6$ (Caldwell {\etal}1991).
\end{enumerate}
These differences in the \hlf\ have usually been interpreted as
resulting from corresponding differences in star formation properties.
For example, the break in slope (Point 1, above) has been suggested to
be caused by a physical transition between normal \hii\ regions and
the class of 
supergiant \hii\ regions such as 30 Doradus in the Large Magellanic Cloud
(KEH).  Likewise, the difference in slope for arm vs. interarm
regions, and among various Hubble types (Points 2 and 3), have been
attributed to differences in gasdynamics and molecular cloud mass
spectrum (\eg Thronson, Rubin, and Ksir 1991; Rand 1992; KEH).
However, we feel that it may be premature to draw a direct connection
between the form of the \hlf\ and details of star forming
environment such as these.  There are other effects that can influence
the form of the \hlf, that have not been adequately explored.

One important effect, in particular, is the influence of simple
evolution in the ionizing clusters, and hence, luminosity of the host
\hii\ regions.  There are several empirical factors that suggest that
nebular luminosity evolution is an important effect in the form of the
\hlf.  The steeper slopes seen in interarm regions of spiral galaxies
(Point 2, above) have been identified by von Hippel \& Bothun (1990)
as evidence of an aging effect in those populations.
Assuming star formation takes place primarily in the spiral density waves
delineated by the arm regions, then the interarm regions should
exhibit a typically older population of nebulae in the wake of this
star formation activity.
In addition, von Hippel \& Bothun cite the behavior of \Ha\ equivalent
widths as  
further compelling evidence of the importance of luminosity evolution.
As a cluster ages, its associated \Ha\ emission decreases, while
that of the underlying red stellar continuum increases, thereby
reducing the \Ha\ equivalent width as a function of time.  They
demonstrate that for the \hii\ regions in NGC 628, there is indeed a
clear decrease in \Ha\ equivalent width with decreasing \Ha\
luminosity, indicating higher numbers of evolved nebulae among the
fainter populations.  Finally, Knapen {\etal}(1993) and Rozas, Knapen,
\& Beckman (1996b) show that the slope
of the nebular luminosity vs. volume relation is slightly flatter in the
interarm regions of five galaxies as compared to the arm regions.  As
emphasized by the authors, this may be caused by observational
selection effects, but if real, we suggest that it is consistent with a higher
tendency for shell-like geometries in the interarm nebulae, as would
be expected for an older population. 

All of these arguments suggest that evolved nebulae will be an
important population in the \hlf.  Von Hippel \& Bothun (1990) carried
out the first detailed investigation of the evolutionary effect on the
\hlf.  They constructed two models to reproduce the observed slope for
NGC 628:  one with zero evolution, in which the power-law slope of the
luminosity function is due entirely to the cluster mass spectrum; and one
in which the distribution in nebular luminosity is due entirely to a
different age and number distribution of objects with the same
initial cluster mass.  The \hlf\ resulting from the evolutionary model
is sensitive to the stellar initial mass function (IMF), since the relative
numbers of stars at a given ionizing luminosity become important when
these stars become the dominant contributor to the nebular luminosity.
Recognizing this, von Hippel \& Bothun oriented their study toward exploring the use of
the \hlf\ to place constraints on the IMF.  As a result, they did not
explicitly examine the consequences of aging on the form of the \hlf\
itself.  In addition, their assumption of uniform initial cluster
masses in the evolutionary model is of only limited use, since actual
initial cluster masses apparently have a power-law distribution (e.g., Oey
\& Clarke 1997; see below).  Therefore, we will here investigate 
the effect of cluster aging on the form of the \hlf\ itself. 

\section{Saturated Population}

We have carried out a preliminary investigation of this problem as
part of a different study (Oey \& Clarke 1997), and
review the important relevant issues here.  A first critical
point is the fact that not all clusters have the
same relative contribution to ionizing luminosity from the various
stellar masses.  Such a situation is true for rich clusters that sample
the IMF well, but breaks down for sparse
clusters in which small-number statistics determine the relative
ionizing contribution of different stellar masses.  We shall refer to
the rich clusters with good stellar statistics as ``saturated'' with
respect to the IMF, and those with poor statistics as ``unsaturated.''
The importance of this fundamental difference in the dominant ionizing
contributors between the saturated and unsaturated clusters is
demonstrated by a break in the predicted slope of the unevolved \hlf.
This change in slope was clearly demonstrated by McKee \& Williams
(1997; hereafter MW97), who performed Monte Carlo simulations of
clusters drawn from a truncated power-law 
distribution in the numbers of member stars $N_*$, and a 
Scalo (1986) IMF.  Since $L\propto N_*$ for the saturated objects,
the \hlf\ slope is the same as the parent slope in
$N_*$; but for unsaturated populations, the \hlf\ slope is
distinctly flatter.  This flattening is caused by the increased
scatter in the $L$'s contributed by a bin of
given $N_*$.  The unsaturated condition occurs when this scatter is
large compared to the mean $L$ at the given $N_*$, hence the slope of
the \hlf\ flattens.  MW97 observed that in their
simulations, this change in slope occurs at $\lsat=\lup$, the luminosity
contributed by a single star at the upper-mass limit $m_{\rm up}$, of
the IMF.  Stellar models currently suggest that 
$\lup$ is in the range $38.0 \lesssim\log \lup\lesssim
38.5$ (Schaerer \& de Koter 1997; Vacca, Garmany, \& Shull 1996;
Panagia 1973).  We caution however, that the slope break $\lsat$ in
the \hlf\ does not necessarily occur at $\lup$, as we show in \S 4.3.  

Investigation of evolutionary effects on the \hlf\ should therefore
consider the differences in stellar ionizing populations between the
saturated and unsaturated regime.
We examined the behavior of the saturated case analytically in Oey \&
Clarke (1997).  We considered two extremes in the formation of
ionizing clusters:  a single-burst scenario, in which all objects are
created at the same time; and a continuous-creation scenario.
In what follows, we will also assume a constant IMF throughout.

For the case of the single burst, it is apparent that, since
all objects follow the same luminosity evolution, the initial slope of
the \hlf\ is simply preserved.

\begin{figure*}
\vspace*{-1.7 truein}
\hspace*{0.8 truein}
\epsfbox{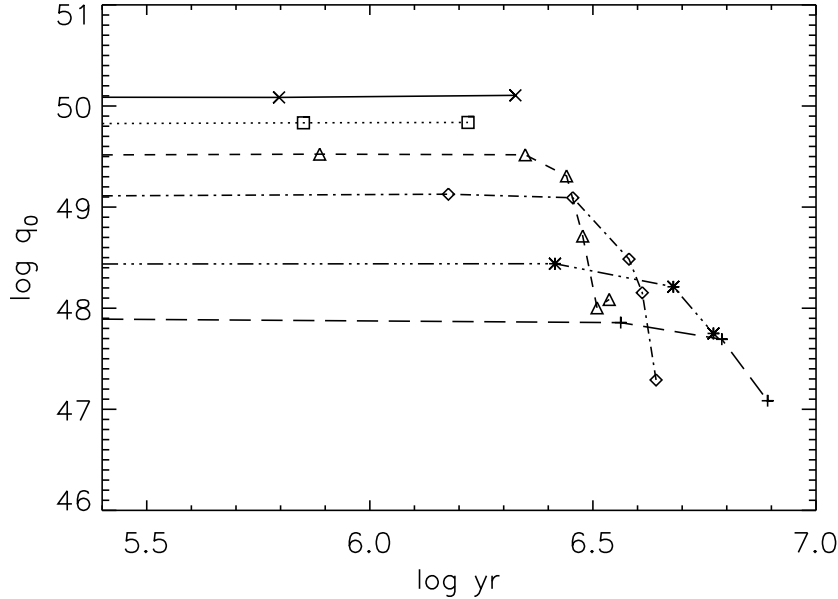}
\vspace*{-10 pt}
\caption{
Evolution of ionizing photon emission rate $q_0$ (sec$^{-1}$) from
Schaerer \& de Koter (1997).  The stellar models are, from bottom, 20,
25, 40, 60, 85, and 120$\msol$.
\label{stellev}}
\end{figure*}

However, in the case of continuous creation, such a conclusion is not
obvious.  We described the luminosity evolution of the
\hii\ regions in terms of a fading function $f(t)$ (Oey \& Clarke 1997):
\begin{equation}
L = L_0\ f(t) \quad ,
\end{equation}
where $L_0$ is the zero-age nebular luminosity.  Stellar population
synthesis work (\eg Leitherer \& Heckman 1995; Beltrametti,
Tenorio-Tagle, \& Yorke 1982) suggests that the tail of the fading
function $f(t)$ is well-described by a power-law in time $t$, so that 
\begin{equation}\label{f}
f = \left\{ \begin{array}{ll}
	1 ~~, & \quad t < t_{\rm ms} \\
	& \\
	\Bigl(t/t_{\rm ms}(\mup)\Bigr)^{-\eta} ~~, & \quad t \geq t_{\rm ms}(\mup)
	\end{array} \right.
\end{equation}
For saturated objects, the onset time, $t_{\rm ms}(\mup)$, for the
fading is the same for all objects.  Nebular fading causes initially
bright objects to 
contribute to lower-$L$ bins at later ages, thus if the epoch of
fading dominates over the epoch of zero-age luminosity, we would
expect the observed \hlf\ to be steeper than the unevolved one.  
We found that for $f(t)$ of the form in equation~\ref{f}, and
reasonable ranges for $m_{\rm up}$, $t_{\rm ms}(\mup)$, 
and $\eta$, that essentially no change in \hlf\ slope is expected.
There is an exception for initially shallow slopes of $ <
\eta^{-1}+1$, for which the observed slope indeed steepens to a value
$a = \eta^{-1}+1$.  This critical value of the slope is probably in the range
1.2 -- 1.4, which is lower than most measured \hlf\ slopes, hence it
is likely that this effect is unimportant.
These effects are described in further detail in Oey \& Clarke (1997).
Thus in the saturated regime, no change in \hlf\ slope
is reasonably expected.

\section{Monte Carlo Models}

The statistical uncertainties that dominate the unsaturated clusters
render an analytic investigation of these objects unfeasible.  This is
due primarily to the fact that many different combinations of stellar
masses can contribute to any given luminosity, and there is no
simple way to relate the total nebular luminosity to the stellar
composition, and hence, luminosity evolution.  We have therefore
modeled the \hlf\ with Monte Carlo simulations to investigate its
behavior in the unsaturated regime.  

Our model draws the integer number of ionizing stars from a power-law
distribution, so that $N(N_*)\ dN_*$ is the number of clusters with
quantities of stars in the range $N_*$ to $N_* + dN_*$:
\begin{equation}\label{N*}
N(N_*)\ dN_* = N_*^{-\beta}\ dN_* \quad .
\end{equation}
We consider ``clusters'' containing a minimum of one star, with the
probability of larger clusters decreasing with $N_*$ according to the
exponent $\beta$.
Our models use the RAN1 portable random number generator described by
Press {\etal}(1986), to generate the random deviate $x$, having values
between 0 and 1.  The distribution in $N_*$ given by equation~\ref{N*}
is then obtained by 
\begin{equation}
N_* = \Bigl[-x \Bigl(1-\beta\Bigr)\Bigr]^{1/(1-\beta)}\quad ,
\end{equation}
where $N_*$ is assigned to the nearest integer.
Since the saturated end of the \hlf\ does not
change slope, as 
outlined in the previous section, we adopt $\beta = 2$ as a 
typical value found for \hlf s in nearby galaxies (\eg KEH).
Unlike MW97, we do not consider
a distribution that is truncated at an upper limit in $N_*$ for our
default models, although this condition is relaxed in \S~3.3.

For each cluster, we then draw $N_*$ stars from a stellar mass
distribution truncated above a mass limit $\mup$:
\begin{equation}\label{imf}
n(m)\ dm\propto \biggl[\Bigl(\frac{m}{\mup}\Bigr)^{-\gamma} - 1\biggl]
	\ dm \quad ,
\end{equation}
where $n(m)\ dm$ is the number of stars in the range $m$ to $m+dm$.
These are generated with an algorithm analogous to that for $N_*$.
Equation~\ref{imf} describes a power-law for $m\ll\mup$, and is the same
analytic form used by MW97 for their $N_*$
distribution and IMF; we refer to this form as a truncated power-law,
following their nomenclature.  We use a Salpeter (1955) power-law
slope of $\gamma = 2.35$, and adopt an upper-mass limit $m_{\rm
up}=100\msol$.  We include stars down to a lower-mass limit $m_{\rm
lo}=17\msol$ (see below).  The \Ha\ luminosity 
$l$ associated with individual stars is dependent on the mass and
main-sequence lifetime $t_{\rm ms}$, for which we also use power law
parameterizations: 
\begin{equation}\label{ml}
m\propto l^\delta \quad ,
\end{equation}
normalized so that $l_{\rm up}\equiv l(m_{\rm up}) = 3\times
10^{38}\ergs$; and
\begin{equation}\label{mti}
m\propto t_{\rm ms}^{-d} \quad ,
\end{equation}
normalized so that $t_{\rm ms}(m_{\rm up}) = 2.8$ Myr (Schaerer
{\etal}1993).  For stars in the range $17 < m < 100\msol$,
we find a value of $d=0.7$ from Schaerer {\etal}(1993).  
We caution the reader that the power-law representations of
equations~\ref{ml} and \ref{mti} are approximations; the relations
steepen significantly toward lower masses.  As discussed below, larger
values of $\delta$ imply a greater sensitivity of $l$ to $m$, thereby
increasing the range of $L$ in the unsaturated regime, for a given
range of $m$.  We adopt $\delta \sim 1.5$, which yields a reasonable
$\log\ l_{\rm lo} \sim 37.3$ for $m_{\rm lo}=17\msol$.  
For illustrative purposes, it is helpful to
show a larger dynamic range of unsaturated objects, though we caution
that in reality the range of these objects is likely to be 
somewhat smaller for our mass limits, since the $m - l$ relation
flattens toward higher masses (\eg Leitherer 1990).
We have adopted the value
of $m_{\rm lo}$ to coincide roughly with the value at which the $l-m$
relation turns over more steeply, since stars below $\sim 17\msol$
do not contribute strongly to the ionizing fluxes.  There is also
a relative increase in $t_{\rm ms}(m)$ at lower masses, but not as
significant as the change in $l(m)$.  It is useful to bear in mind
that the representative values of $\delta$ and $d$ will increase
toward lower masses.


We consider two different formulations for the stellar ionizing
rate $q_0$ with time.  The first simply approximates that
$q_0$ remains constant until $t_{\rm ms}$ and is zero thereafter.
The second formulation accounts for evolution of $q_0$ during the
stellar main-sequence phase.  Figure~\ref{stellev}
shows $q_0$ for stars of different initial masses as a
function of time, from the models of Schaerer \& de Koter (1997).
These models suggest that $l$ remains fairly constant until a
time after which it decreases roughly along the relation,
\begin{equation}\label{stelleveq}
l = \lup\ \biggl(\frac{t}{t_{\rm ms}(m_{\rm up})}\biggr)^{-\zeta} \quad ,
\end{equation}
where $\zeta\sim 5$, and $t$ is the age from birth.  Given the
uncertainties in massive star evolution and ionizing fluxes, we adopt
the crude approximation of constant $l$ until the time of intersection
with equation~\ref{stelleveq}, after which we adopt $l$ from that
relation.  One consequence of this formulation is that stars with
initially different luminosities can have the same $l$ during certain of their
fading periods.  Figure~\ref{stellev} suggests that this may not be
far from reality.  We caution that the assumption of coeval star
formation within individual nebulae is likely to be an
oversimplification, and that the stellar age spread is typically
similar to the main-sequence lifetime of the most massive stars (\eg
Massey {\etal}1995).  Therefore, the early nebular luminosity evolution 
could potentially increase somewhat before the onset of fading.  
Specific modeling of star formation scenarios would be necessary to
investigate this effect.

Our adopted lower-mass cutoff introduces artificial fluctuations 
that can be seen in all our models at low $L$.  The quantized nature
of the stellar ionizing sources, combined with the truncation 
of stellar masses below $m_{\rm lo}$, causes artificial kinks in the
distribution of nebular luminosities at low $L$.  For example, the
most prominent of these features appears at $\log L = 37.6$ in our
models, or 0.3 dex above the lower-$L$ cutoff of $\log l_{\rm lo} =
37.3$.  This dip results from the fact that all nebulae at $L\lesssim
37.6$ contain single stars, selected according to the IMF, whereas
two-star nebulae begin to contribute above this luminosity. 
To more clearly demonstrate the features in the \hlf, we
therefore include $10^4$ clusters in our simulations, although this
number is higher than is seen in most nearby galaxies.  Finally, we
truncate our model \hlf s below the minimum luminosity $l_{\rm lo}$,
which is that due to a single star of mass $m_{\rm lo}$.  Although
objects originating at higher $L$ evolve to luminosities below $l_{\rm
lo}$ for cases incorporating main-sequence evolution, the omission of
objects originating below $l_{\rm lo}$ renders the 
model populations incomplete below that value.

\subsection{Single Burst}

\begin{figure*}
\vspace*{-0.8 truein}
\hspace*{-0.25 truein}
\epsfbox{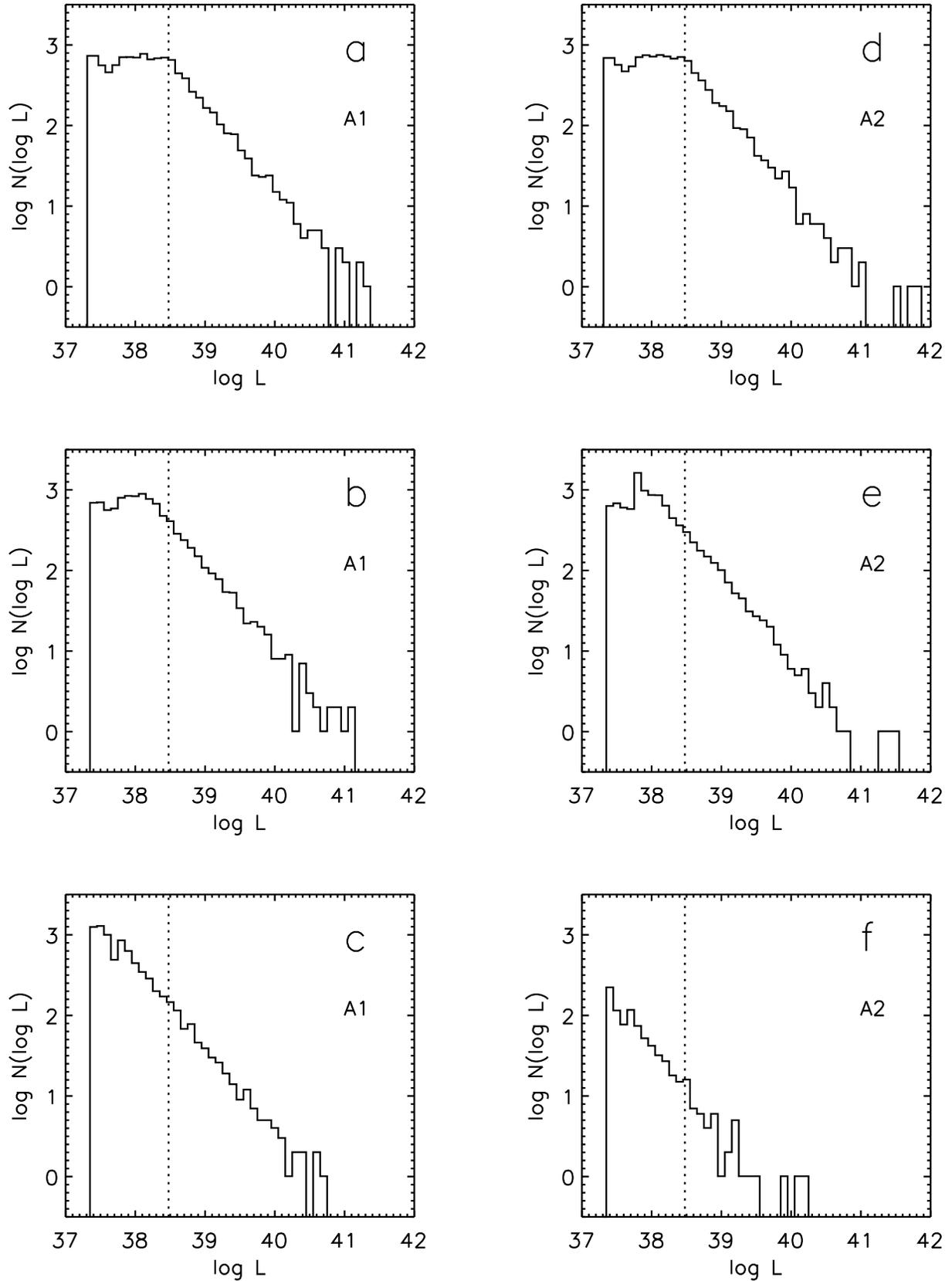}
\caption{Model \hlf s for a single burst.  Model A1 (panels $a-c$)
considers constant $l$ over main sequence lifetimes; Model A2 (panels
$d-f$) considers main sequence fading of $l$ as described in
\S 3.  The dotted line marks $\lsat=3\times 10^{38}\ \ergs$.  Ages of 0,
4, and 7 Myr are shown from top to bottom, respectively.  The dip
near $\log L \sim 37.6$ is a quantization artifact induced by the
lower-mass cutoff.  [Note that the slope of $\log N(\log L)$ 
vs. $\log L$ is $1-\beta$.]
\label{unsat}}
\end{figure*}

Figure~\ref{unsat} shows two models for the evolution of an \hii\
region population created in a single burst, with $\delta = 1.5$,
$d=0.7$, and $17<m<100 \msol$.  The left column of Figure~\ref{unsat}
(panels $a-c$) shows model A1, with constant $l$ during the main-sequence;
while the right column (panels $d-f$) shows model A2, with $l(t)$ 
as described above, to include main-sequence fading.  The zero-age
\hlf s are given in the top panels, with the middle and bottom panels
showing ages of 4 and 7 Myr, respectively.  Note that the two zero-age
models (top panels) are statistically identical.  To aid comparison of
the models, the dotted line corresponding to $\lsat = 3\times 10^{38}\
\ergs$ is shown for reference.

\begin{figure*}
\vspace*{-0.25 truein}
\hspace*{-0.25 truein}
\epsfbox{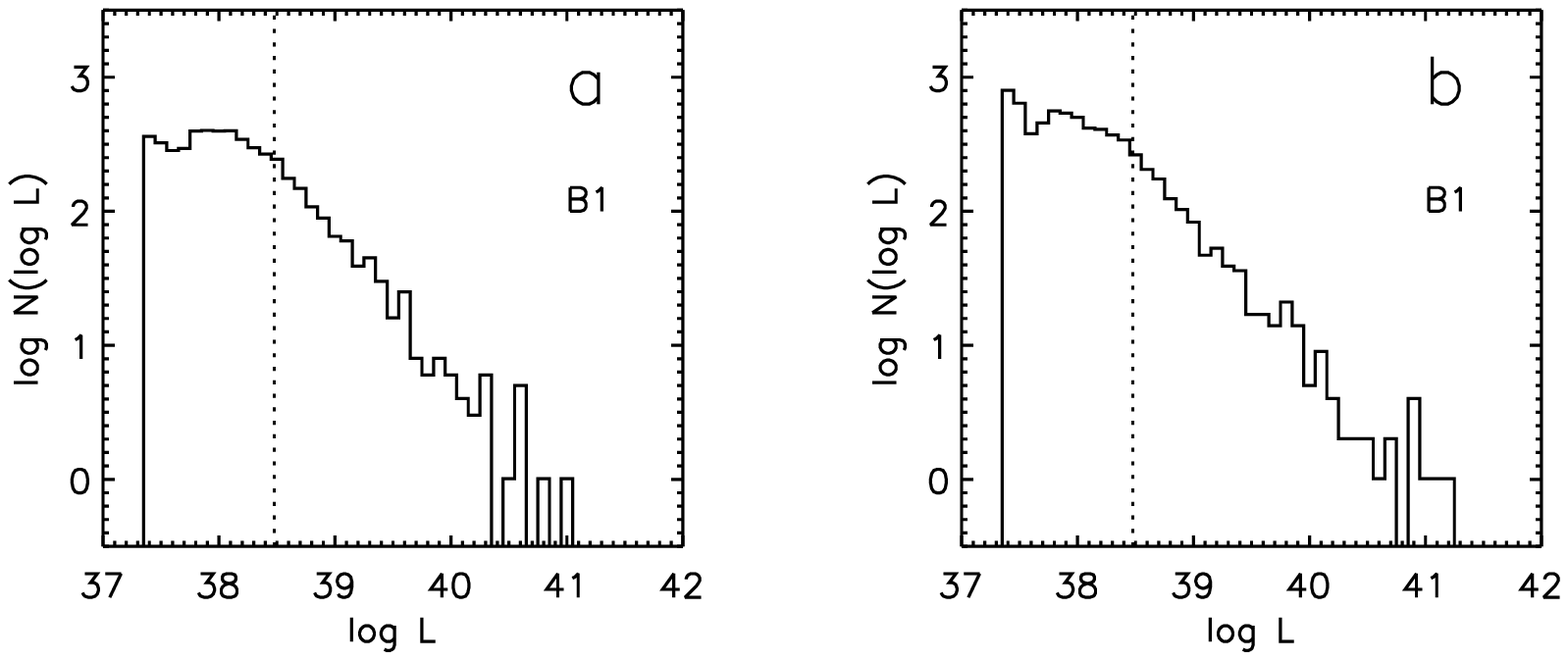}
\vspace*{-6.25 truein}
\caption{\hlf\ Model B1, for continuous creation, with
constant $l$ over main sequence lifetimes.  Panel $a$ shows
an age of 6 Myr after the onset of continuous creation, before
steady-state is achieved.  Panel $b$ shows the steady-state form. 
(The dip near $\log\ L\sim 37.6$ is the same quantization artifact seen
in Figure~\ref{unsat}.)
\label{unsatcc}}
\end{figure*}

As found in the simulations of MW97, the zero-age
models in Figure~\ref{unsat} (top panels) show a flattening of the
\hlf\ slope below a saturation luminosity.  For the adopted stellar
parameters, this transition from saturated to unsaturated populations
occurs at $\log\lsat=\log\lup=38.5$.  Equation~\ref{ml} shows
that the extent of this unsaturated regime is determined 
by $\delta$:  a larger $\delta$ implies a greater
sensitivity of $l$ to $m$, thereby increasing the
range in $L$ of this population, for a given
range of $m$.  This effect also causes the slope in the 
unsaturated regime ($L < \lsat$) to be dependent on $\delta$ because
the scatter in $L$ contributed by individual stars is larger for large
$\delta$, and this scatter determines the flattening in slope.  Hence,
larger values of $\delta$ will induce a flatter slope.  As mentioned
above, our adopted value of $\delta=1.5$ is an
overestimate for the more massive stars in the IMF, hence the
unsaturated slope should be somewhat steeper near $\lsat$ than
is seen in our models.

\hii\ regions with $L > \lsat$ are not sensitive to the $m - l$
relation since statistical effects are unimportant in this regime.
The slope of the \hlf\ in the saturated regime is $\beta$, the same as
that of the parent distribution in $N_*$ (equation~\ref{N*}), as
described in \S 2. 

Figures~\ref{unsat}$b$ and $c$ show Model A1 at ages of 4
and 7 Myr respectively.  These ages should not be regarded too
literally, since the evolutionary state at a given age is dependent on
equation~\ref{mti}, which is an approximate parameterization, as well
as equations~\ref{imf} and \ref{ml}.  For reference, $t_{\rm
ms}(m_{\rm lo})=9.7$ Myr for our adopted $d=0.7$.  It is immediately
apparent that the inflection in slope 
moves toward progressively lower luminosities with age.  This is
primarily a reflection of the group evolution of the saturated
population to lower luminosities; the entire distribution at $L>\lsat$ in
Figure~\ref{unsat}$a$ is moving toward lower luminosities in
Figures~\ref{unsat}$b$ and $c$.
For cases such as this, in which the zero-age $\lsat=\lup$, the
slope inflection at later ages now corresponds to $l$ of the most
massive star still present in these objects.  More generally, the rate
at which the saturated objects fade in $L$ has a dependence of
$t^{\delta/d}$, as determined by equations~\ref{ml} and \ref{mti}.


Model A2 includes main-sequence evolution of the ionizing fluxes.  We
caution that the models including the main-sequence evolution should
be interpreted qualitatively, owing to the crude parameterization
of the stellar evolution.  The principal feature is that Model A2
clearly evolves much faster than Model A1.  The other important effect
is that noticeable, sharp structure in the \hlf\ becomes apparent near the
transition from saturated to unsaturated objects, as seen in
Figure~\ref{unsat}$e$ as contrasted to Figure~\ref{unsat}$b$.  This
results from the fact that 
clusters dominated by a few, high-mass stars are now capable of evolving
to lower $L$ than clusters dominated by many, unevolved, low-mass stars.
In other words, stellar evolution allows the low-$L$ end of the
saturated population to overtake, toward lower luminosities, the
high-$L$ end of the unsaturated population.  The details of this
structure in the \hlf\ are dependent on the details of the 
main-sequence luminosity evolution. 

\subsection{Continuous Creation}

Figure~\ref{unsatcc} shows simulations for a population where the
\hii\ regions are being continuously created (Model B1).  This model is the
same as that in the single burst case with no main-sequence
luminosity evolution (Model A1), but now with each new cluster
formed at a uniform incremental time after its predecessor.  The total
distribution of initial luminosities, $\log L_0$, of all the objects,
which represents the parent, unevolved \hlf, is the same as the
zero-age distributions in Models~A1 and A2 (Figures~\ref{unsat}$a$ and
$d$).  Figure~\ref{unsatcc}$a$ shows the 
distribution at an age of 6 Myr after the onset of continuous creation, but
before the population has achieved its steady-state form.  The
steady-state \hlf, which occurs at $t > t_{\rm ms}(m_{\rm lo})$, is shown
in the Figure~\ref{unsatcc}$b$.

We may consider the \hlf\ for continous creation to be a sum of 
individual bursts with a continuous age distribution.  Since the
single burst \hlf\ evolves predominantly as a progression of the
saturated objects, with slope $\beta$, to the lower luminosities of
the unsaturated regime, we therefore expect that the
slope in the region $L<\lsat$ will be intermediate between
its flatter value in the unevolved case and $\beta$.
Figure~\ref{unsatcc}$b$ shows that this is indeed the case.  We can
also see in Figure~\ref{unsatcc}$a$, that the intermediate
evolutionary stage shows 
the region of steepening moving in from higher luminosities, 
since the low-luminosity bins are not yet affected by saturated,
evolved objects.  We also generated a Model B2, incorporating main sequence
luminosity evolution, analogous to Model A2.  The steady-state form of
that model also shows an intermediate-value slope at $L <
\lsat$, but slightly flatter than its analog in
Model B1 because of the structure present at $L<\lsat$
in the burst case (Model A2).

\subsection{Unsaturated Population}

\begin{figure*}
\vspace*{-2.0 truein}
\hspace*{-0.25 truein}
\epsfbox{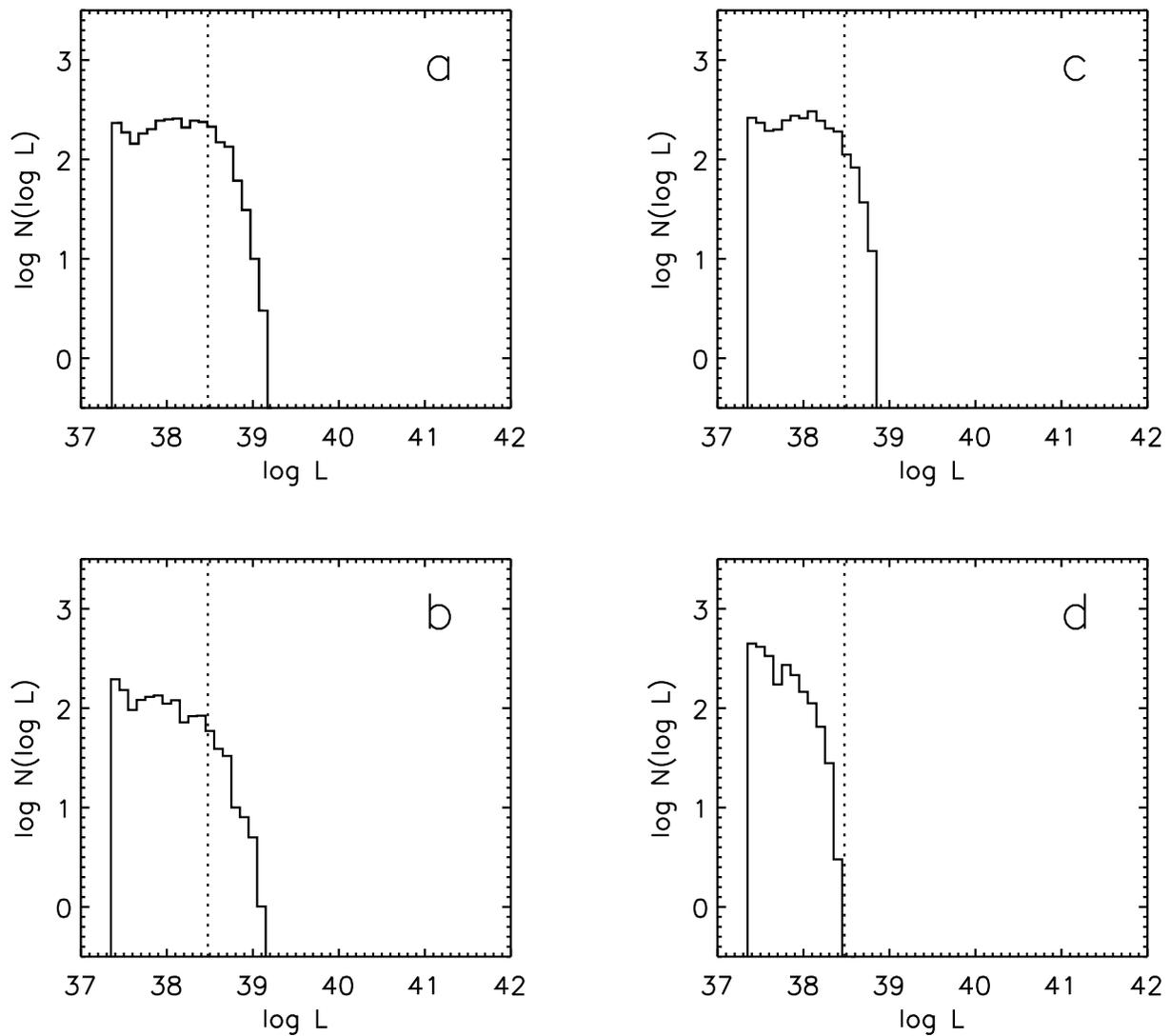}
\vspace*{-3.1 truein}
\caption{Models with the same parameters as in Figures~\ref{unsat}
and \ref{unsatcc}, but for populations with $N_*\leq 10$. $(a)$
Unevolved population; $(b)$ steady-state, continuous creation; $(c)$
single burst at age 4 Myr; $(d)$ single burst at age 7 Myr.  To aid
comparison, these models are plotted on the same scale as
Figures~\ref{unsat} and \ref{unsatcc}. 
\label{unsatonly}}
\end{figure*}

In Figure~\ref{unsatonly}, we show models for a population of only
unsaturated clusters.  These are drawn from the same power law
distribution in $N_*$ (equation~\ref{N*}) as the previous models, but
only for $N_* \leq 10$.  For our default stellar parameters with
$17<m<100 \msol$, the saturation transition in $N_*$ is roughly
between 10 and 20 stars.  We use the same stellar parameters as in the
previous models, now with a total of 3000 clusters.
The left side of Figure~\ref{unsatonly} shows the simulation for 
the unevolved population ($a$), and continuous creation ($b$);
while the right side shows the single burst case at ages 4 Myr ($c$)
and 7 Myr ($d$).  These models do not include main sequence
luminosity evolution, whose consequences are analogous to those
discussed for Models A2 and B2 above.

It is apparent in Figure~\ref{unsatonly} that the primary difference
in the \hlf\ between this population and the complete power-law
sampling in $N_*$, is that the high-$L$ tail of the unsaturated
objects falls off very steeply.
The highest-luminosity, unsaturated clusters result from a chance
assemblage containing a disproportionate number of high-mass
stars.  Therefore, the
distribution of these objects must fall more
steeply than that of a saturated population, owing largely to the
scarcity of the high-mass stars.  The actual slope in this
regime is once again determined by the detailed statistics of the
cluster compositions, and equations~\ref{imf} and \ref{ml}.  

At $L<\lsat$, the behavior of these unsaturated populations 
is identical to that seen in the models with a full distribution in $N_*$, 
since, as discussed above, this range in $L$ is
dominated by the unsaturated objects.  
It is only at the high-$L$ limit itself, in the evolved case of
Figure~\ref{unsatonly}$d$, that the sudden drop due to
sheer lack of objects, effectively generates a steep slope.
Note that at luminosities below this high-$L$ drop-off, the \hlf\
in this model recovers a slope of $\beta$, analogous to
Figure~\ref{unsat}$c$. 

\section{Observed \hlf s}

The features and behavior found in the simulated 
\hlf s can provide a useful interpretation of the patterns found in
observed \hlf s.

\subsection{Arm and Interarm \hlf s}

\begin{figure*}
\vspace*{-0.8 truein}
\epsfbox{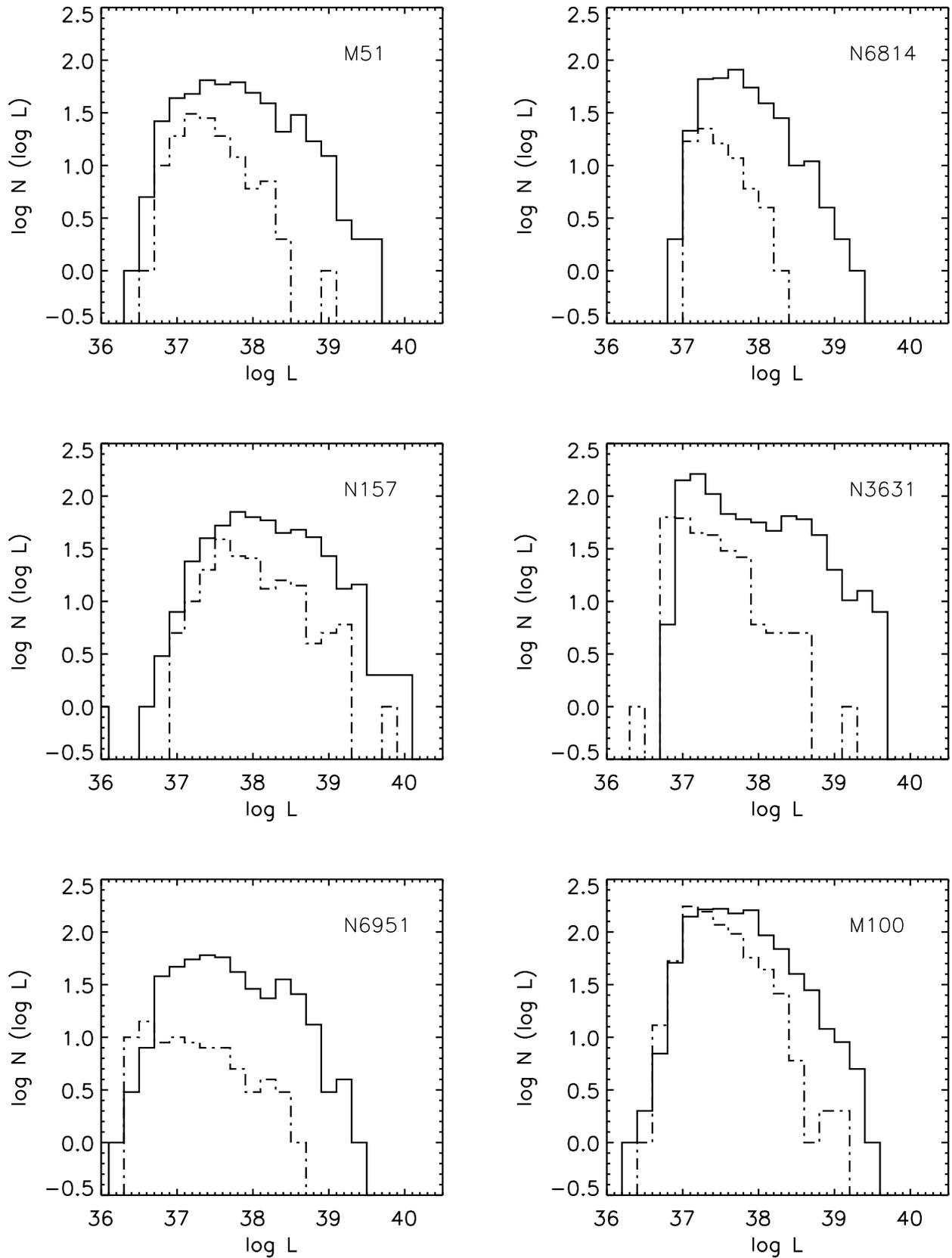}
\caption{Observed \hlf s for arm (solid lines) and interarm
(dot-dashed lines) populations in grand design spirals:  M51 (Rand
1992); NGC 6814 (Knapen {\etal}1993); NGC 157, NGC 3631, and NGC 6951
(Rozas {\etal}1996a); M100 (Knapen 1997).
\label{arms}}
\end{figure*}

In Figure~\ref{arms}, we reproduce the published \hlf s for arm and
interarm populations in six grand design spirals.  The solid
histograms show the arm \hlf s, and the dot-dashed lines show the
interarm \hlf s.  In each case, the peak of the interarm
\hlf\ occurs at a fainter luminosity than that for the arm regions,
suggesting that the interarm \hlf\ mirrors that of the arms at
lower $L$.  This shift in peak occurs despite the smaller total
numbers of \hii\ regions in the interarm areas.  Owing to
incompleteness effects, we might otherwise expect the smaller numbers
to induce a turnover at, if anything, higher $L$.  We do caution
that blending of nebulae (Deharveng et al. 1988; KEH) in the more
crowded arm populations could also introduce a flattening in the
observed \hlf\ at lower $L$.  However, as pointed 
out previously (Rand 1992; Knapen {\etal}1993; KEH), the highest-$L$
objects in the interarm regions are apparently always fainter than
those in the arms.  This is again consistent with a group shift in
luminosity. 

These features exhibit the behavior of our model \hlf s
for the burst scenario.  This is consistent with the expectation that
the arm nebulae are largely coeval, and are undergoing a current burst of
cluster formation; whereas the interarm regions represent, on average,
an older population remaining from previous star formation.  As
seen in Figure~\ref{unsat}, the inflection in 
the \hlf\ occurs at progressively lower values with age.  
Note that even if incompleteness affects the location of the peak for
the arm objects, which we take to reflect the unevolved population,
the relative shift in the peak for the interarm objects is still
consistent with evolution, owing to the steeper slope at lower $L$.
This is because the low-$L$ bins are more robust to incompleteness for a
steep slope than for a shallow one, since the proportion of missing
objects is the same in each case.  Therefore,
while the absolute location, or existence, of the \hlf\ peak may be 
suspect due to observational limitations, the observed shift in peak
for the interarm regions is fully consistent with our model because a
steeper slope is expected at low $L$ (cf. Figure~\ref{unsat}$d$ and
$f$).   Indeed, given that the  
\hlf\ peak is often substantially lower in $\log N (\log L)$ for the interarm
regions, this suggests that the actual peak of the distribution has
moved to luminosities much lower than the empirical completeness
limit, as is illustrated, for example, in Figure~\ref{unsat}$f$.

The aging effect can also explain the steeper \hlf\ slopes that
are sometimes reported for the interarm regions as compared 
to the arm regions of spiral galaxies (Point~2 in \S 1).
If the actual nebular luminosity function conforms to the two-slope
shape seen in the models, and if the two slopes are not fitted separately, 
then the net effect is that the measured slope will be
intermediate between the two.  In our single burst model, the location
of the break moves to lower $L$.  Therefore,
over a fixed range of $L$, the contribution of the shallower slope to
the composite, measured \hlf\ slope will be less for an evolved
burst population, than for an unevolved one.  As is apparent in
Figure~\ref{arms}, it is difficult to pinpoint the location of the
slope change.  We mention once again that our simulations show
a flatter slope in the unsaturated regime than would be expected,
owing to the parameterization of equation~\ref{ml}.
Furthermore, the observed location of $\lsat$ will
also vary with galactic internal extinction, which most \hlf s,
including the ones in Figure~\ref{arms}, do not take into account.
With these difficulties, we suggest that the slopes of the interarm regions
do appear to be the same as those in the saturated regime of the arm
populations.

Thus the differences observed in the \hlf s for arm
and interarm regions are not necessarily due to changes in star
formation environment, but are likely to be affected by evolution.
The time scale between the passage of spiral density waves is of order
$\sim40$ Myr (\eg Rand 1993) at any given location in the disk, which is
much longer than the observable life expectancy of the \hii\ regions.
Thus, the spatial and luminosity distribution of 
nebulae between the arms does imply that there is at least a certain
amount of bona fide interarm star formation taking place.  However,
based on the arguments presented in \S 1, we feel that an aging 
population is likely to be an important component in the interarm
regions, and that the behavior of the \hlf\ is consistent with this
interpretation.  With the present data and the observational
uncertainties mentioned above, combined with uncertainties in
predicted $q_0$ and its evolution, we have not attempted to place
meaningful age constraints on the evolved population.  However, it will
be interesting to re-examine the data more closely, and search for an
evolutionary progression in the \hlf\ with distance behind the leading
edge of the spiral arms.  We hope to carry out such an investigation
in the near future.

\subsection{The Hubble Sequence}

The correlation of earlier type galaxies with steeper slopes (Point 3 in
\S 1) can be explained by systematically lower cutoffs in $N_*$ in
those clusters.  KEH have shown compelling evidence that
these galaxies have fewer stars per cluster, since
the mean nebular luminosities in the early type spirals
are substantially lower than those in late types, despite the larger
total numbers of \hii\ regions in the earlier types.  Therefore,
unless the stellar ionizing population varies drastically, it is
probable that the 
size of the parent clusters must themselves be smaller on average in
the earlier type galaxies.  The same conclusion has been found for Sa
galaxies (Caldwell {\etal}1991), although the total numbers of \hii\
regions are again smaller in those galaxies.

\begin{figure*}
\vspace*{-0.25 truein}
\hspace*{-0.25 truein}
\epsfbox{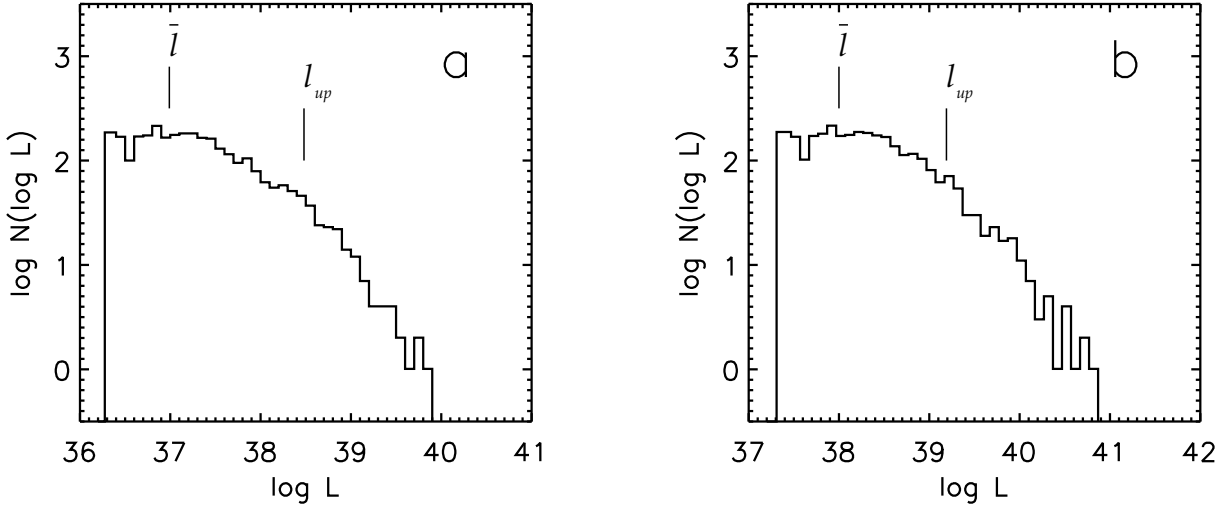}
\vspace*{-6.25 truein}
\caption{($a$)  Model for $3<m<100 \msol$; and ($b$)  model for
$17<m<300 \msol$.  The location of $\bar{l}$ and $\lup$ are indicated
in each case.
\label{trans}}
\end{figure*}

The decrease in typical $N_*$ per cluster with Hubble type (KEH) could
easily result from a decrease in the upper cutoff of the $N_*$ power-law.
The earliest type galaxies would have populations composed only of
unsaturated objects, as would correspond to Sa galaxies,
where the maximum $\log L$ is typically in the range $38.0 - 38.5$
(Caldwell {\etal}1991).  This is the same range in values as the
estimates for $\lup$, as mentioned earlier, hence it is indeed the regime of
unsaturated objects.  As seen in \S 3.3, the models that
describe this population show a steeper high-luminosity tail above
$\lsat$, than do the populations with saturated objects.  Indeed, the
slopes in Sa galaxies have been found to be even steeper than those
for Sb -- Sc types:  the typical \hlf\ slope $a\sim2.6$ for Sa
galaxies (Caldwell {\etal}1991), as compared to $a\sim 2.0$ for Sb --
Sc galaxies (KEH; Banfi {\etal}1993).  The model in
Figure~\ref{unsatonly}$b$, for continuous creation of unsaturated objects,
would therefore be the most appropriate of our simulations to describe
the Sa \hlf s.  The form of the Sa \hlf s shown in
Figure~5 of Caldwell {\etal}(1991) show a consistent resemblance with
our models in Figure~\ref{unsatonly}.  Thus, ironically, although
these nebulae are dominated by unsaturated objects whose
characteristic slope is flatter than the saturated case, the high-$L$ 
drop-off will induce a steeper measured slope in these populations.

The Sb -- Sc galaxies would then show a slope intermediate between the
Sa and Im types (cf. Figures~4$b$ and 3$b$), as the upper cutoff in
$N_*$ approaches the exclusively 
unsaturated regime.  Indeed, KEH suggest that the \hlf\ in Sb -- Sc
galaxies is truncated at an upper-$L$ limit.  This is quantitatively
supported by MW97, who also find evidence for a
truncated \hlf\ in the Milky Way and Sb--Sc galaxies, but not in
irregular galaxies.  Since this limit is still
at $L$ above the regime of unsaturated objects, these findings are
consistent with our interpretation of a progressively lower
upper-$N_*$ cutoff with earlier Hubble type.  There may be exceptions
caused by temporarily enhanced star formation, in which large numbers
of saturated clusters are formed.  This scenario can explain the
larger range in \hlf\ slope seen in earlier type galaxies (KEH).
Indeed, a slope steepening at high $L$ reported in some earlier type
galaxies could well correspond to a cutoff; indeed, KEH note that it
is often impossible to distinguish such a turnover from a cutoff.
Thus the steepening in slope for disk galaxies across the Hubble
sequence can be explained by a simple trend in upper cutoff of the
$N_*$ distribution, and need not invoke a change in the parent
slope $\beta$ of $N_*$.

\subsection{The Slope Transition}

In \S 3, we denoted the luminosity of the slope break in the
initial \hlf\ by $\lsat$ and noted that, as in the models of MW97,
this often corresponds to the luminosity
contributed by most luminous star in the IMF, $\lup$.  We
however caution that this correspondence is dependent on the stellar
parameters.  As an example,  
Figure~\ref{trans} exhibits the \hlf\ resulting from two models with
identical IMF slope and $m-l$ relation, but with different stellar
mass limits.  Figure~\ref{trans}$a$, with $3<m<100 \msol$, exhibits an \hlf\
that more closely resembles in form the simulation by MW97.
There are two principal slope breaks apparent:  one near
$\lup$; and one near $\bar{l}$, below which the slope appears quite flat.  
The existence of a minimum cluster membership number, $N_*=1$, causes
this flattening, or in some cases even positive slope, for
luminosities below the expectation value of $L$ containing single
stars, i.e., the average stellar luminosity $\bar{l}$.  
At $\bar{l}<L<\lup$, there is an intermediate slope in the \hlf\ that
joins the flat regime to the saturated regime of slope $\beta$.
A small transition region of $\lesssim 0.3$ dex is noticeable in this
intermediate-slope region near both $\bar{l}$ and $\lup$, which
results from the form of the IMF adopted in equation~\ref{imf}.  With
instead a sharp IMF truncation at $\mup$, the slope breaks would occur cleanly
at $\bar{l}$ and $\lup$.

One can now see that certain stellar mass ranges and values of $\delta$
can change the relative location of the two breaks such that one or
the other could disappear.  The model in Figure~\ref{trans}$b$, with
$17<m<300 \msol$, shows an example where the $\lup$ slope break is
virtually undetectable at $\log \lup=39.2$.  The smaller difference
between $\bar{l}$ and ${\lup}$ has compressed the intermediate-slope
range in $L$-space, thereby steepening its slope to a value closer to
$\beta$, which represents the maximum slope possible in this regime.
An observed \hlf\ similar to Figure~\ref{trans}$b$ would
most likely be measured to have only one slope in the regime $L>\bar{l}$.
Note that the models in Figure~\ref{trans} are probably 
unrealistic, since $\delta=1.5$ is not representative over such large
mass ranges, and we only show these for illustrative purposes.
However, these examples demonstrate that breaks in observed \hlf s
should be interpreted with caution, and may only provide a lower limit
to the most luminous star in the IMF.
Furthermore, the tendency of the highest-mass stars to approach
similar ionizing emission rates $q_0$ (\eg Leitherer 1990), compounds the
insensitivity of $\lup$ to $\mup$.  We therefore conclude that,
unfortunately, the \hlf\ is not a useful probe of $\mup$.

However, the insensitivity of $\lup$ to $\mup$ suggests that
the high-$L$ end of the \hlf\ should be a reliable indicator of
$\beta$ for saturated nebular populations, since it diminishes the
likelihood of a varying $\mup$ 
as an explanation for the observed variations in the \hlf.  
The principal caveat is that a truncation at maximum $N_*$ not be
confused with a steep slope of $\beta$, as may have been done for Sa
galaxies in the past (see above).
It is also apparent that, since the form of the \hlf\ is sensitive to
the stellar parameters, that demonstrable variations in the
unsaturated regime of the zero-age or continuous creation \hlf\ 
could possibly indicate substantial variations in the stellar
parameters of the hottest stars.  


We note that stochastic variations in $\mup$ within a galaxy will not
change the location of the break if the distribution of $\mup$ is
gaussian.  We confirmed this by drawing $\mup$ from a parent normal
distribution of $\sigma = 15\msol$ and the same mean value of
100$\msol$.  The resulting \hlf\ is indistinguishable from that with a
constant $\mup$.

Empirically, a flattening at lower
$L$ is often seen in many galaxies (Point 1 in \S 1), but at varying critical
luminosities.  While the models for a single burst show that the
transition point is not necessarily fixed, $\lsat$ cannot be greater
than $\lup.$ 
At present, the value of $\lup$, the \Ha\ luminosity associated with
$\mup$, is not well-determined.  Our models have adopted a value of
$\lup = 3 \times 10^{38}\ \ergs$, which is uncertain within a factor
of a few, and is a high estimate.  Most published \hlf s are not
corrected for internal extinction within the galaxies, which 
will affect the location of the break.  As can be seen in Figure~8$b$
of Walterbos \& Braun (1992), extinction correction for the
highly-inclined case of M31 can shift the location of the apparent
turnover by
an order of magnitude.  It is also important to distinguish between
a real slope change and artificial slope flattening induced by
limitations such as incompleteness, and the more insidious problem of
object blending, mentioned above, for more distant galaxies.
In view of the difficulty in pinpointing the location of a slope
change, and the variations induced by reddening and possible aging
effects, we suggest that the published \hlf\ slopes are thus far
consistent with our models.

We note that KEH modeled the lower-$L$ end of the \hlf\ by
combining the $m - l$ relation with the IMF for single-star
\hii\ regions.  This representation is somewhat different from ours,
since we include multiple-star, low-$L$ objects that will contribute
to the same luminosity bins occupied by bright, single-star nebulae.
Our slope in this regime will therefore tend to be slightly flatter
(more positive) than theirs.  This will affect estimates for the
fraction of total galactic ionizing emission due to faint \hii\
regions below empirical completeness limits, such that those using the
KEH model are an upper limit for similar stellar parameters.

Additional models of the \hlf\ by Feinstein (1997) came to our
attention during the refereeing process.  These models consider only
nebular populations with luminosities scaled from objects with a
saturated IMF, but with $m_{\rm lo}$ extending down to $5\msol$.  We
emphasize that, since unsaturated clusters are not considered, the
slope turnover seen in those models is therefore 
{\it not} caused by the same mechanism seen in our work, or 
that of KEH and MW97.  Rather, the slope break seen in Feinstein's
models result from the adopted lower cutoff in the $L$ distribution,
combined with an extended creation period.  The zero-age, saturated
\hlf\ is described by the power-law $\beta$, which simply shifts to
lower luminosity with advancing age, as exhibited by the saturated
tail of our models in Figure~\ref{unsat}.  Feinstein's continuous
creation models are therefore a superposition of many power-laws with
a continuous range of ages.  The lower-$L$ cutoff seen in those models
therefore induces a flattening in the \hlf\ below that luminosity.
This is best illustrated in the right panel of Feinstein's Figure~3,
which varies the lower-$L$ cutoff according to assumed total stellar
mass, for continuous creation.  The slope break occurs at 
the lower-$L$ cutoff, hence, as pointed out by the author, 
this parameter determines the location of the break.  We caution,
however, that the lower-$L$ cutoffs used in Feinstein's models are
necessarily artificial, owing to the consideration of only saturated
\hii\ regions.  Therefore, the slope breaks seen in these models must
be regarded as artificial to the same extent.  If, however, nebular
creation scenarios exist with such a strong lower-$L$ cutoff,
Feinstein's models demonstrate another mechanism which can also induce
a slope break.

Previous work has sometimes suggested an empirical slope change
at high luminosities of $\log L\sim 38.6$ or even higher (Rozas
{\etal}1996a; Rand 1992; KEH).  If a break at these high luminosities
is real, it is unlikely to result from the stochastic effect below $\lsat$.
However, if the high-$L$ slope is steep enough, it could well be a
manifestation of the maximum $N_*$ cutoff in those galaxies, which is
expected to vary in the scenario we presented in \S 4.2.  It could
also result from a change in $\lsat$, although this appears unlikely,
as argued above.  But factors other than those that we have considered
can affect the form of the \hlf\ as well.  For example, Rozas
{\etal}(1996a,b) suggest that such a break at high $L$ results from 
a transition between radiation-bounded nebulae at fainter
luminosities, to density-bounded objects at brighter luminosities.
Another potentially 
important effect is a differential extinction pattern among the nebular
population, and of course, the usual differences in IMF slope and star
forming conditions.  Given the currently available data, the
systematic presence of an additional, high-luminosity slope change 
at a fixed $L$ remains controversial.  

\section{Conclusion}

We have modeled the \hlf\ with Monte Carlo simulations that draw $N_*$
for each cluster from a power-law distribution of constant slope, and
for each cluster, $m$ from a Salpeter IMF truncated at
$\mup=100\msol$.  As found by MW97, this produces
shallower slopes corresponding to the population dominated by 
unsaturated objects, and a steeper, high-$L$ side dominated by
saturated ones.  The saturated slope is $\beta$, the slope of the
$N_*$ distribution (equation~\ref{N*}).  The transition 
luminosity $\lsat$ in the unevolved \hlf\ may correspond to $\lup$, the 
\Ha\ luminosity due to a single star of mass $\mup$, but this depends
on the assumed stellar parameters.  The observed \hlf s do often show
flatter slopes at lower luminosities that are consistent with the
transition at $\lsat$ (\eg Walterbos \& Braun 1992; Scowen, Dufour, \&
Hester 1992).

We have hitherto not discussed the form of the \hlf\ below $l_{\rm
lo}$, which is unexplored in our models.  Since the slope $\delta$ of
the $m - l$ relation (equation~\ref{ml}) steepens, the \hlf\ will
continue to grow flatter for the same $N_*$ power law, extending to
lower luminosities.  At $\log L 
\lesssim 36$, contamination from planetary nebulae may become
important (Walterbos \& Braun 1992), while confusion and background
diffuse \Ha\ emission will undoubtedly limit the \hlf\ in this regime.


Our simulations show that for a single-burst case, the slope transition
moves to lower $L$ as the population ages and the saturated objects
grow fainter (Figure~\ref{unsat}).  Hence, the 
steady-state model for continuous creation shows a slope in the
region $L<\lsat$ that is intermediate between its value in the unevolved
\hlf\ and the slope $\beta$ of the saturated objects
(Figure~\ref{unsatcc}).  Models that 
include main-sequence evolution for stellar ionizing luminosities show
a faster evolution than those that do not, with a noticeable peak
around the transition between saturated and unsaturated populations
(Figure~\ref{unsat}$e$).
The steady-state model for continuous creation
incorporating stellar evolution yields a slightly shallower slope than
that for the case without stellar evolution, but still steeper than
the unevolved case.  Models for a population of only
unsaturated clusters with $N_*\leq 10$ show a much steeper drop-off at
$L> \lsat$ (Figure~\ref{unsatonly}), where the \hlf\ is normally
dominated by saturated objects.  The behavior at $L<\lsat$ is generally
the same as that for the models with fully populated $N_*$, with the
exception of a high-$L$ drop-off in an evolving burst.

The patterns seen in observed \hlf s in nearby galaxies can be
explained by the behavior of these models.  If \hii\ regions in spiral
arms of galaxies represent a current burst of coeval star formation,
then the steeper slopes seen in interarm vs. arm regions can be
explained as an aging burst compared to a current burst population.
Since single slope measurements will yield a composite of the
saturated and unsaturated slopes, the zero-age \hlf\ will show a
flatter slope owing to the larger range of unsaturated objects.  The
lower peak $L$ and maximum $L$ seen in the interarm regions are also consistent
with a dominant contribution from an aging population.  

The steeper slopes seen in earlier galactic Hubble types can be
explained by the lower cutoff in maximum $N_*$.  As seen in the models for
populations of only unsaturated clusters (Figure~\ref{unsatonly}), the
high-$L$ tail is much steeper than in populations with higher $N_*$.
A truncation at the high end of the \hlf\ has been suggested by
KEH and MW97.  
Since the mean $N_*$ does apparently decrease in earlier type galaxies
to the values in unsaturated clusters (KEH; Caldwell {\etal}1991), the
steeper \hlf\ slopes are likely to simply result from a lower maximum $N_*$.
A slope dropoff therefore may also be seen at $L$ corresponding to the
maximum $N_*$. 

Thus, we suggest that the current data are consistent
with evolutionary effects and the upper-$N_*$ cutoff being the primary
determinants of the shape of the \hlf.  
Indeed, it may be unnecessary to invoke
any other parameters for the nebular and cluster properties to
explain the behavior of the \hlf\ as described in \S 1.
If this is true, it is rather remarkable, for example, that grand design
spirals with strong star formation induced by tidal interactions, would
show the same power-law distribution in $N_*$ as those found in normal
disk and irregular galaxies.  However, we
again acknowledge that other, unexplored effects, such as 
gas density bounding, differential
reddening, and variations in IMF and other stellar parameters, can
influence the form of the \hlf.  Nor can we rule out suggested
environmental influence like variations in slope of the molecular
cloud mass spectrum.  But thus far, it appears that the power-law
slope $\beta$ of $N_*$ might vary even less than was suggested by the
observed slopes of the \hlf, even though the rate and total star formation
can vary by one to two orders of magnitude.  Further detailed
observations of the \hlf, and re-examination of those in the
literature, are of great interest in evaluating 
the possibility of a universal $\beta\sim 2$.  Suggestively, Caldwell
{\etal}(1991) find that the luminosity function of knots seen in the
$B$ continuum of four of their Sa galaxies do show such a power-law
exponent $\sim 2$. 

As is done with the luminosity functions of distant galaxies,
it may thus be possible to use the form of the \hlf\ to understand
features in the most recent star formation history ($\lesssim 10$ Myr)
of their host galaxies, since 
individual bursts can leave their imprint on the \hlf.  Our models,
for example, show that an aged burst population is distinguishable
from continuous creation of low-$N_*$ clusters
(cf. Figures~\ref{unsat}$c$,$f$ and \ref{unsatonly}$b$).
Feinstein (1997) also demonstrates effects of different creation
histories.  At present, 
it is difficult to make useful quantitative constraints, owing to present
constraints on stellar and cluster parameters, but it is at least
possible gain qualitative insight on the recent star formation history.


\acknowledgments

We thank Rob Kennicutt, Johan Knapen, and the referee, Carlos
Feinstein, for comments on the manuscript.  We also thank
Johan Knapen for access to data for M100 in advance of publication,
and John Beckman for an interesting discussion.


\begin{thebibliography}{}

\bibitem{} Banfi M., Rampazzo R., Chincarini G., \& Henry R. B. C.,
	1993, AA, 280, 373
\bibitem{} Beltrametti M., Tenorio-Tagle G., Yorke H.W. 1982,
	AA, 112, 1
\bibitem{} Caldwell N., Kennicutt R., Phillips A.C., \& Schommer R.A.,
	1991, ApJ 370, 526
\bibitem{} Deharveng L., Caplan J., Lequeux J., Azzopardi M.,
	Breysacher J., Tarenghi M., \& Westerlund B., 1988, AAS, 73, 407
\bibitem{} Feinstein C., 1997, ApJS 112, 29
\bibitem{} Kennicutt R. C., Edgar B. K., \& Hodge P. W., 1989, ApJ,
	337, 761 (KEH)
\bibitem{} Knapen J. H., 1997, MNRAS, submitted
\bibitem{} Knapen J. H., Arnth-Jensen N., Cepa J., \& Beckman J. E.,
	1993, AJ, 106, 56
\bibitem{} Leitherer C., 1990, ApJS, 73, 1
\bibitem{} Leitherer C. \& Heckman T. M. 1995, ApJS, 96, 9
\bibitem{} Massey P., Lang C. C., DeGioia-Eastwood K., \& Garmany
	C. D. 1995, ApJ, 438, 188
\bibitem{} McKee C. F. \& Williams J. P. 1997, ApJ, 476, 144 (MW97)
\bibitem{} Oey M. S. \& Clarke C. J. 1997, MNRAS, 289, 570
\bibitem{} Panagia N. 1973, AJ, 78, 929
\bibitem{} Press W. H., Flannery B. P., Teukolsky S. A., \& Vetterling
	W. T., 1986, {\sl Numerical Recipes}, Cambridge:  Cambridge
	Univ. Press, 195
\bibitem{} Rand R. J., 1992, AJ, 103, 815
\bibitem{} Rand R. J., 1993, ApJ, 410, 68
\bibitem{} Rozas M., Beckman J. E., \& Knapen J. H. 1996a, AA, 307, 735
\bibitem{} Rozas M., Knapen J. H., \& Beckman J. E. 1996b, AA, 312, 275
\bibitem{} Salpeter E. E. 1955, ApJ, 121, 161
\bibitem{} Scalo J., 1986, Fund. Cosmic Phys., 11, 1
\bibitem{} Schaerer D., de Koter A. 1997, AA, 322, 598
\bibitem{} Schaerer D., Meynet G., Maeder A., \& Schaller G., 1993,
	AAS, 98, 523
\bibitem{} Scowen, P. A., Dufour, R. J., \& Hester, J. J. 1992, AJ,
	104, 92
\bibitem{} Thronson H. A., Rubin H., \& Ksir A., 1991, MNRAS, 252, 550
\bibitem{} Vacca W. D., Garmany C. D., \& Shull J. M. 1996, ApJ, 460,
	914
\bibitem{} von Hippel T. \& Bothun G., 1990, AJ, 100, 403
\bibitem{} Walterbos R. A. M. \& Braun R. 1992, ApJS, 92, 625

\end{thebibliography}
\end{document}